\documentclass[aps,prl,preprint,showpacs,floatfix]{revtex4}
\usepackage[dvips]{graphicx}
\usepackage{subfigure}
\begin{document}

\title{Non-dispersive optics using storage of light}

\author{Leon Karpa}
\author{Martin Weitz}
\affiliation{Institut f\"ur Angewandte Physik der Universit\"at Bonn, Wegelerstr. 8, 53115 Bonn, Germany}

\date{\today}

\begin{abstract}
We demonstrate the non-dispersive deflection of an optical beam in a Stern-Gerlach magnetic field. An optical pulse is initially stored as a spin-wave coherence in thermal rubidium vapour. An inhomogeneous magnetic field imprints a phase gradient onto the spin wave, which upon reacceleration of the optical pulse leads to an angular deflection of the retrieved beam. We show that the obtained beam deflection is non-dispersive, i.e. its magnitude is independent of the incident optical frequency. Compared to a Stern-Gerlach experiment carried out with propagating light under the conditions of electromagnetically induced transparency, the estimated suppression of the chromatic aberration reaches 10 orders of magnitude.
\end{abstract}

\maketitle

Electromagnetically induced transparency (EIT) is a quantum interference effect that allows to render otherwise opaque atomic media transparent \cite{10.1103/RevModPhys.77.633(eg)}. Destructive interference of absorption amplitudes here allows for a suppression of optical absorption. Media prepared under the conditions of electromagnetically induced transparency have interesting properties, as an extreme reduction of the optical group velocity \cite{Nature.397.594-598(1999),PhysRevLett.82.5229,PhysRevLett.83.1767} and allow for applications e.g. in the fields of metrology \cite{PhysRevLett.69.1360, Katsoprinakis2006, Kominis2003}, non-linear optics \cite{PhysRevLett.82.5229, Harris1999} and quantum information science \cite{Duan2001, Chaneliere2005, 10.1038/nature04327, Duan2001a}. More recently, spatially resolved EIT has been demonstrated, and in two experiments the storage of images has been reported \cite{Shuker2008, Vudyasetu2008}. This has prospects in the context of all-optical transmission and processing of arbitrary images. We have recently shown that slow light can be spatially deflected by a Stern-Gerlach magnetic field, yielding evidence for a non-zero effective magnetic moment of the dark polariton, the quasiparticle physically associated with slow-light propagation \cite{2006NatPh.2.332K}. Due to the extremely spectrally sharp variation of the group velocity, the beam deflection is highly dispersive \cite{Sautenkov2007}, i.e. when the light is slowed down by say a factor two, as is possible by variation of the signal beam frequency by a few kHz in the spectrally sharp EIT transparency window, the deflection doubles. In another experiment, the deflection of an optical beam has been achieved by means of phase imprinting from a spatially inhomogeneous off-resonant optical field during light storage \cite{Schnorrberger2009}.

Here we demonstrate the non-dispersive deflection of an optical beam traversing a medium under the conditions of electromagnetically induced transparency. Our experiment is based on a light pulse stored as a spin wave in thermal rubidium vapour, onto which a phase gradient is imprinted by a spatially inhomogeneous Stern-Gerlach magnetic field. The observed beam deflection upon reacceleration of the optical pulse is found to be independent of the signal beam frequency, thus suppressing chromatic aberrations. More generally, we provide a proof of principle experiment demonstrating that the dynamic deceleration and acceleration of light possible under the conditions of electromagnetically induced transparency allows to surpass limitations of conventional optics. We also wish to point out that the obtained beam deflection cannot be understood in terms of Fermat's principle of the shortest optical path.

Before proceeding we note that the reduction of chromatic and spherical aberrations has a long history in the field of both light and matter wave optics. For example, state of the art optical objectives employ high order corrections to reduce chromatic aberrations over the entire visible spectrum \cite{Bass1995}. While those optical elements rely on light traversing a spatial sequence of optical elements with different dispersional properties (different glass materials), our approach uses temporal variations of the optical group velocity in an atomic gas and phase imprinting.
\begin{figure}[ht]
    \begin{center}
        \includegraphics[height=4cm]{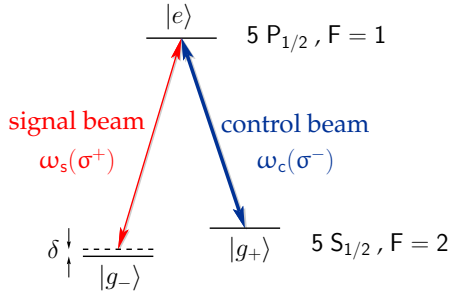}
            \caption{\label{fig:figure1} Simplified scheme of the relevant atomic levels.
            A $\sigma^-$-polarized control field and a $\sigma^+$-polarized
            signal field with optical frequencies $\omega_c$ and $\omega_s$ couple the levels
            $\left|g_{+}\right\rangle$, $\left|e\right\rangle$ and $\left|g_{-}\right\rangle$,
            $\left|e\right\rangle$ respectively. The two-photon detuning is given by
            $\delta=\omega_s-\omega_c-2g_F\mu_B B_z$.}
        \end{center}
\end{figure}
Our experiment can be understood qualitatively by considering an ensemble of atoms with a $\Lambda$-type coupling scheme, where $\left|g_-\right\rangle$ and $\left|g_+\right\rangle$ denote two stable ground states with Zeeman quantum numbers differing by two and $\left|e\right\rangle$ a spontaneously decaying state, see Fig. \ref{fig:figure1}. A schematic of our experiment is shown in Fig. \ref{fig:figure2a}. The atoms are irradiated by two copropagating optical fields, a weak "signal" field with frequency $\omega_s$ and a stronger "control" field with frequency $\omega_c$ coupling the levels $\left|g_-\right\rangle$, $\left|e\right\rangle$ and $\left|g_+\right\rangle$, $\left|e\right\rangle$ respectively. The magnetic bias field is oriented along the beam axis. Storage of light is performed in a standard way \cite{Phillips2001,Nature.409.490-493(2001)} by adiabatically reducing the intensity of the control beam to zero, so that a signal beam pulse is first slowed and then coherently stored as a spin-wave coherence in the atomic medium. During the light storage, the phase of the spin wave coherence oscillates with the corresponding atomic eigenfrequency. The atoms are subject to a transverse magnetic field gradient, so that the eigenfrequency is not spatially homogeneous over the signal beam profile. Within a storage time $\tau$, the imprinted accumulated phase shift in the Zeeman ground state coherence is
\begin{equation}
\Delta\varphi(x)=\frac{1}{\hbar}\int\limits_0^\tau \Delta E(x)dt
\label{eqn:PhaseGeneral}
\end{equation}
where $x$ denotes the position along the magnetic field gradient. In our experiment both the magnetic bias field $B_z$ and the incident beams are collinear (along the z-axis), so that the differential Zeeman splitting between the ground states $\left|g_+\right\rangle$ and $\left|g_-\right\rangle$ is  $\Delta E(x) = 2 g_F \mu_B B_z(x)$, with $g_F$ as the hyperfine g-Factor and the Bohr magneton $\mu_B$. In the presence of the field gradient, the accumulated phase shift $\Delta\varphi(x)$ is spatially inhomogeneous, which will lead to an optical path length variation $\Delta s(x) = \lambda \Delta\varphi(x)$ upon reacceleration of the signal pulse that is dependent on the transverse position $x$. This is similar as in the directed emission of phased-array antennas in the microwave regime \cite{ArtechhouseBoston1994}. For a constant magnetic field gradient $\frac{dB_z}{dx}$, the associated angle beam deflection $\alpha_{store}$ is easily determined by setting $\Delta\varphi(x) = 2 g_F \mu_B \left( \frac{dB_z}{dx} \right) x \tau / \hbar \cong \alpha_{store} k x$ for small deflection angles. For a magnetic gradient constant in time over a storage period $\tau$, we arrive at
\begin{equation}
\alpha_{store}=\left( \frac{dB_z}{dx} \right) \frac{2 g_F \mu_B \lambda \tau}{h}.
\label{eqn:AlphaStore}
\end{equation}
In our experiment, the magnetic field gradient is active both during the propagation time of the optical signal beam pulse through the dark state medium and the stored light phase. In first order (we continue to assume small deflection angles), the total deflection is expected to be the sum of the two contributions $\alpha_{tot} = \alpha_{store} + \alpha_{move}$, where $\alpha_{move}$ is the Stern-Gerlach angle deflection of a propagating slow light pulse acquired during the phases before and after the storage period of the pulse. In a mechanical model, this contribution to the total deflection can be determined by $\alpha_{move} \cong \frac{\Delta k}{k}$, where $\Delta k = \frac{1}{\hbar}\int\limits^{L/v_g}_0 F_{SG}dt$ with the Stern-Gerlach force  $F_{SG} = \mu_{pol}\left( \frac{dB_z}{dx} \right)$ on an atom-light polariton with effective magnetic moment $\mu_{pol}$,  and L denotes the cell length \cite{2006NatPh.2.332K}. For a group velocity $v_g \ll c$ we have  $\mu_{pol} \cong 2 g_F \mu_B$, and arrive at
\begin{equation}
\alpha_{move} \cong \left( \frac{dB_z}{dx} \right) \frac{2 g_F \mu_B \lambda L}{h v_g},
\label{eqn:AlphaMove}
\end{equation}
where for the sake of simplicity we have set the group velocity $v_g$ to be constant during the course of the deceleration and retrieval procedure. Note that formally also Eq. \ref{eqn:AlphaStore} can be obtained in this ``mechanical'' model, if we assume that the stored light polaritons cannot be moved in the presence of the Stern-Gerlach field, but rather accumulate a change in wavevector $\Delta k$ during the stored light phase that changes the propagation direction of the regenerated beam after retrieval.

It is instructive to compare the variations of the deflection angle in the incident optical frequency. Whereas the variation of the deflection angle acquired during the stored light phase resembles that of an optical grating and is as low as $\frac{\Delta \nu}{\nu} \cong 10^{-10}$ in the spectrally narrow transmission width of an EIT resonance, e.g. $\Delta \nu \cong 20$ kHz in our experiment. In contrast, the optical group velocity varies by a factor of order unity over the spectral width of the dark resonance, so that the chromatic aberration of the Stern-Gerlach deflection of stored light is suppressed with respect to that acquired in the propagating light phase by an expected factor $\frac{\Delta \nu}{\nu}$, i.e. ten orders of magnitude!

In our proof of principle experiment, we do not pulse the magnetic field gradient so that both contributions to beam deflection can be studied, and experimentally compared. Our setup is a modified version of an earlier described apparatus \cite{2006NatPh.2.332K, Karpa2009}. A 50 mm long cell containing thermal $^{87}$Rb-vapor and 10 torr
of neon buffer gas is placed inside a magnetically shielded
region, in which a field coil generates the (longitudinal)
magnetic bias field.
\begin{figure}[ht]
        \subfigure[]{
           \label{fig:figure2a}
           \includegraphics[height=3.5cm]{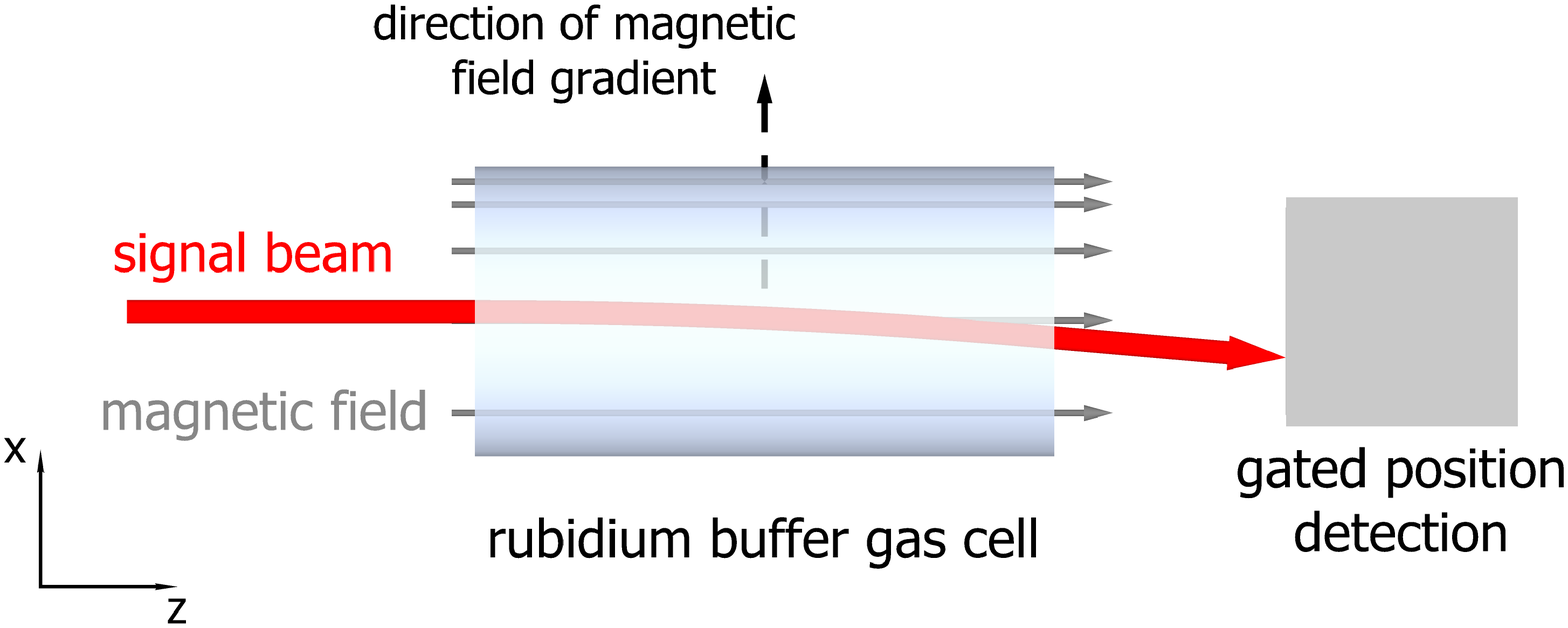}}
    \subfigure[]{
           \label{fig:figure2b}
           \includegraphics[height=3cm]{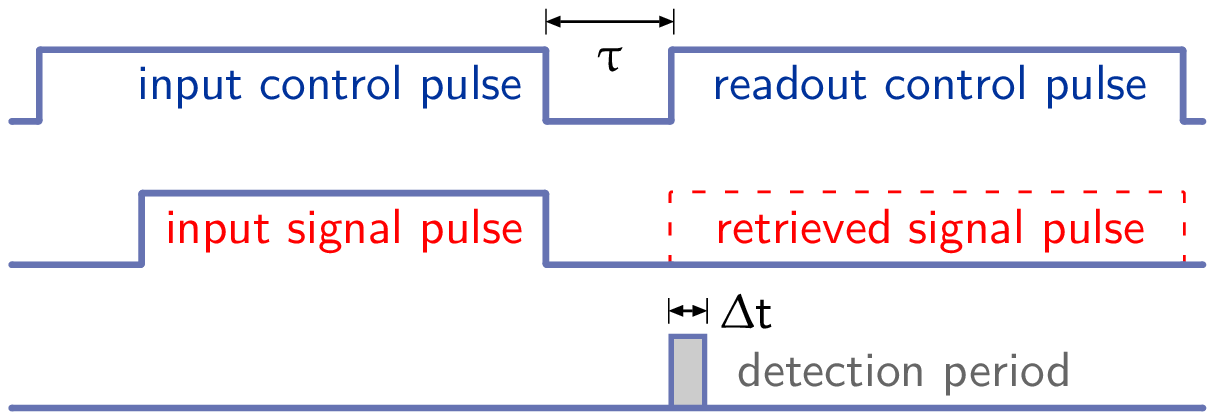}}
     \caption{\textbf{(a)} Schematic diagram of the experiment.
     An optical signal beam is stored as a spin-wave in rubidium
     vapour, and subsequently reaccelerated to finite group
     velocities. A transverse Stern-Gerlach magnetic field gradient causes an
     angle deflection of the released optical beam. \textbf{(b)} Schematic pulse sequence used for the
           experiment. The group velocity of the optical signal
           beam is steered by the intensity of the control beam,
           allowing for storage of the signal field in the medium for a temporal
           period $\tau$ and its subsequent retrieval. Position detection of the
           signal field occurs only in the readout phase after storage for
           a temporal period $\Delta t$, during which an acousto-optic modulator
           placed before a position detecting CCD-camera is activated (not shown
           in Fig. \ref{fig:figure2a}).}

     \label{fig:SchemeAndPulses}
\end{figure}
To produce the desired transverse magnetic
field gradient, a $\mu$-metal strip oriented parallel to the
propagation axis of the optical fields is mounted near the cell,
resulting in a lowering of the magnetic field at the cell side
near the strip. The buffer gas cell apparatus is heated to
temperatures about 85$^{\circ}$C. Both signal and control optical
fields are derived from the same grating stabilized laser diode
source locked to the $F=2\longrightarrow F'=1$ hyperfine component
of the rubidium D1-line near 795 nm. The emitted beam is split
into two paths and guided through separate acousto-optical
modulators (AOMs) to enable controlled variations of the
two-photon detuning and of the individual intensities of the
control and signal fields respectively. The beams are spatially
overlapped and, after being fed through a polarization maintaining
optical fibre, expanded to a 2 mm beam diameter and with opposite
circular polarizations sent to the rubidium cell. The used beam
powers upon entry in the cell were typically 60 $\mu W$ and 330
$\mu W$ for the signal and control beams respectively. In the
limit that the signal beam intensity being always much less than
the intensity of the control beam, most of the relevant population
is in the $F=2, m_F=-2$ sub-level of the 5 S$_{1/2}$ ground state.
In this case the levels $\left|g_-\right\rangle$ and
$\left|g_+\right\rangle$ of the simplified level scheme of Fig.
\ref{fig:figure1} correspond to the $m_F=-2$ and $m_F=0$ Zeeman
sub-levels of the 5 S$_{1/2}, F = 2$
hyperfine ground state component respectively.

After traversing the rubidium cell, the position of the deflected
signal beam is monitored. As the CCD-camera used for position
monitoring does not have sufficient temporal resolution to allow
for a selective detection of the signal beam during readout time
of the storage of light sequence, an acousto-optical modulator is
placed before the camera. The modulator is driven with a
radiofrequency gating pulse of typically $\Delta t = 5$ $\mu s$
length during retrieval of the signal beam, so that position
detection is active only in this temporal window. The position of
the retrieved signal beam pulse is determined by mapping the
intensity distribution onto a CCD camera and calculating the
center of intensity position
of the digitized image.

In initial experiments, we have mapped out the magnetic field
gradient by recording dark resonance spectra with the driving
optical beams passing the experiment cell at different transverse
positions. In these preparatory measurements the signal beam
intensity was monitored by a photodiode. From a series of recorded
spectra, the field gradient was determined to be (1.21 $\pm$
0.07)$\times 10^{-3}$ G/mm for a magnetic bias field of 240 mG.

Subsequently, we moved to a storage of light sequence, and
monitored the deflection of the reaccelerated beam. A schematic of
the used experimental pulse sequence is shown in Fig. \ref{fig:figure2b}. Storage
of light is achieved by irradiating the cell with rectangularly
shaped signal and control beam pulses of 130 $\mu s$ and 150 $\mu
s$ duration respectively, which were switched off adiabatically
such that the falling edge of the signal pulse temporally
coincided with the falling edge of the control pulse. This leaves
the signal pulse stored as a spin-wave in the atomic medium, with
the magnetic field gradient imprinting an accumulated, spatially
inhomogeneous phase shift onto the atomic spin coherence. After a
storage period of variable duration, the coupling laser is turned
back on and the stored coherence is reaccelerated into a
propagating optical field. The regenerated atomic dipoles have
acquired a transversal phase gradient $\Delta\varphi(x)$, and we
expect an angle deflection of the regained optical beam.
\begin{figure}[ht]
  \begin{center}
    \includegraphics [height=4cm]{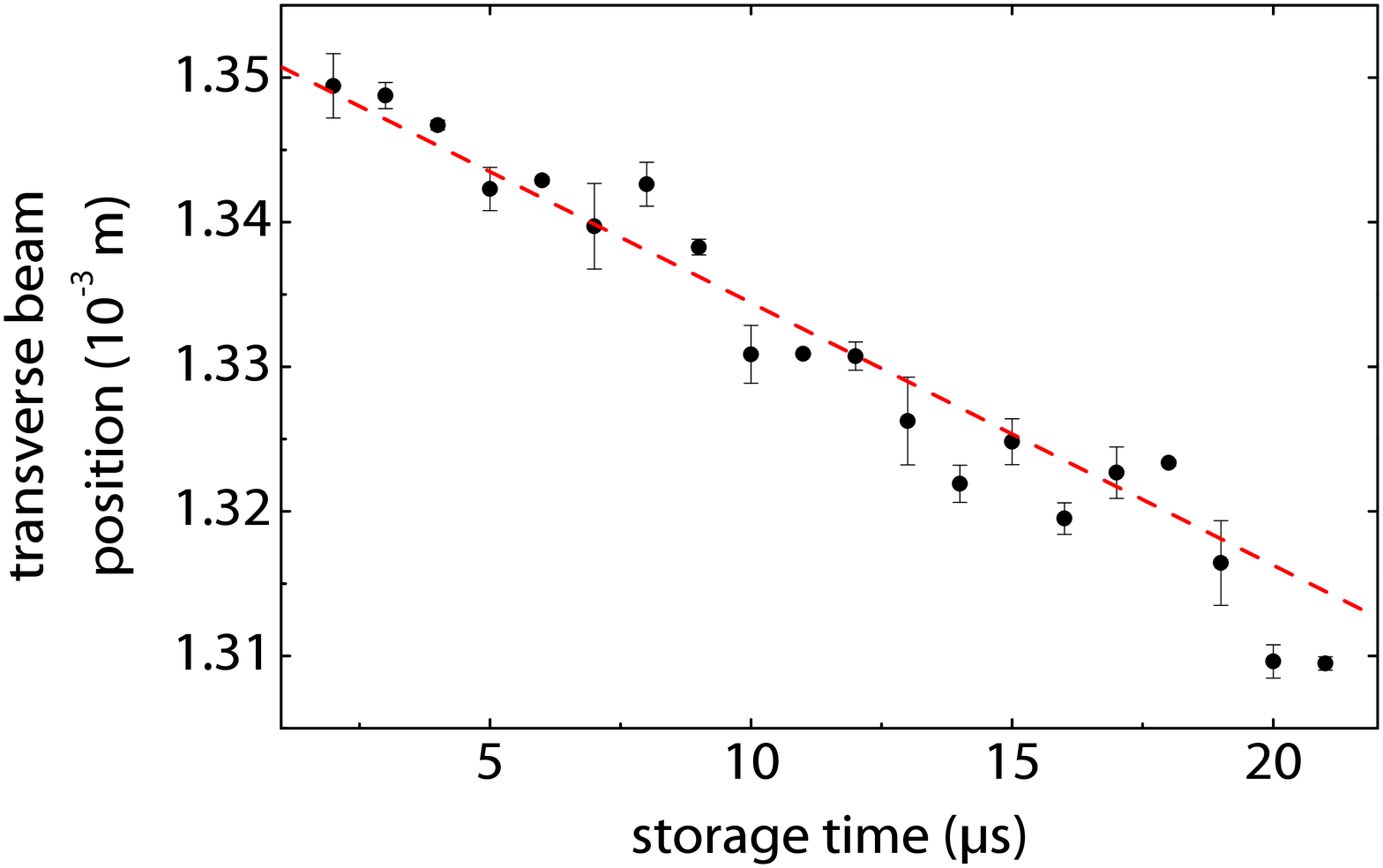}
  \caption{\label{fig:figure3} Measured signal beam position after retrieval as a function
  of the storage time. The data points were recorded with a two-photon
  detuning of $\delta\cong 0$ in the beam center. The points give
  the
  average position values determined in four measurement sessions,
  each acquiring the average of the results of 100 subsequent position
  measurements. The shown error bars were determined from the
  standard deviation of the results of the independent sessions.
  The data points have been fitted with
  a linear function, as shown by the dashed red line.}
  \end{center}
\end{figure}
Fig. \ref{fig:figure3} shows the result of a position measurement
of the retrieved signal beam recorded for different storage period
durations. The measurement gives the signal beam position on our
CCD camera placed at a distance of 1.7 m behind the rubidium cell.
In between the cell and the camera a 1:1 optical telescope was
located consisting of two 200 mm focal length lenses used for
focusing of the signal beam into the acoustooptic modulator
required for temporal gating of the retrieved signal beam pulse, with the
first lens 0.6 m apart from the cell. The experimental data can be
well fitted assuming the expected linear dependence of the
deflection induced by phase imprinting on the storage time (see
Eq. \ref{eqn:AlphaStore}). To verify that the phase imprinting indeed causes an
angular deflection of the reaccelerated beam, after the telescope
an additional lens (f=300 mm focal length) was inserted at the end
of the signal beam path, with its focal plane being congruent with
the detection plane of the CCD camera. In this configuration, only
angle changes cause a change of the camera position signal. An
angle deflection $\alpha$ by the cell at given storage time leads
to a variation of the beam position on the camera $\Delta x \cong
f \alpha$ for small angles.
\begin{figure}[ht]
     \centering
        \subfigure[]{
           \label{fig:figure4a}
           \includegraphics[height=4cm]{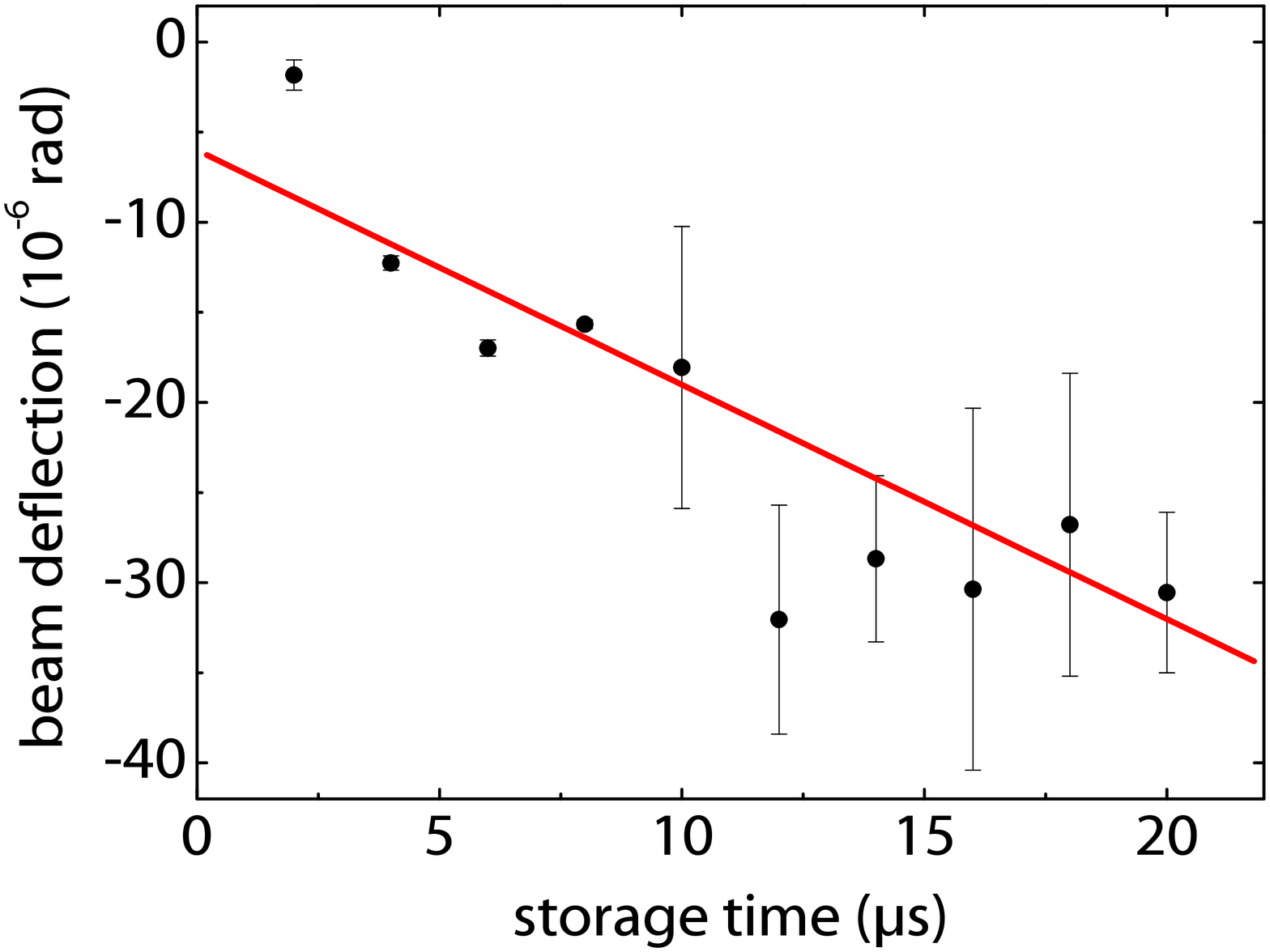}}
\hspace{0.5cm}
        \centering
     \subfigure[]{
           \label{fig:figure4b}
           \includegraphics[height=4cm]{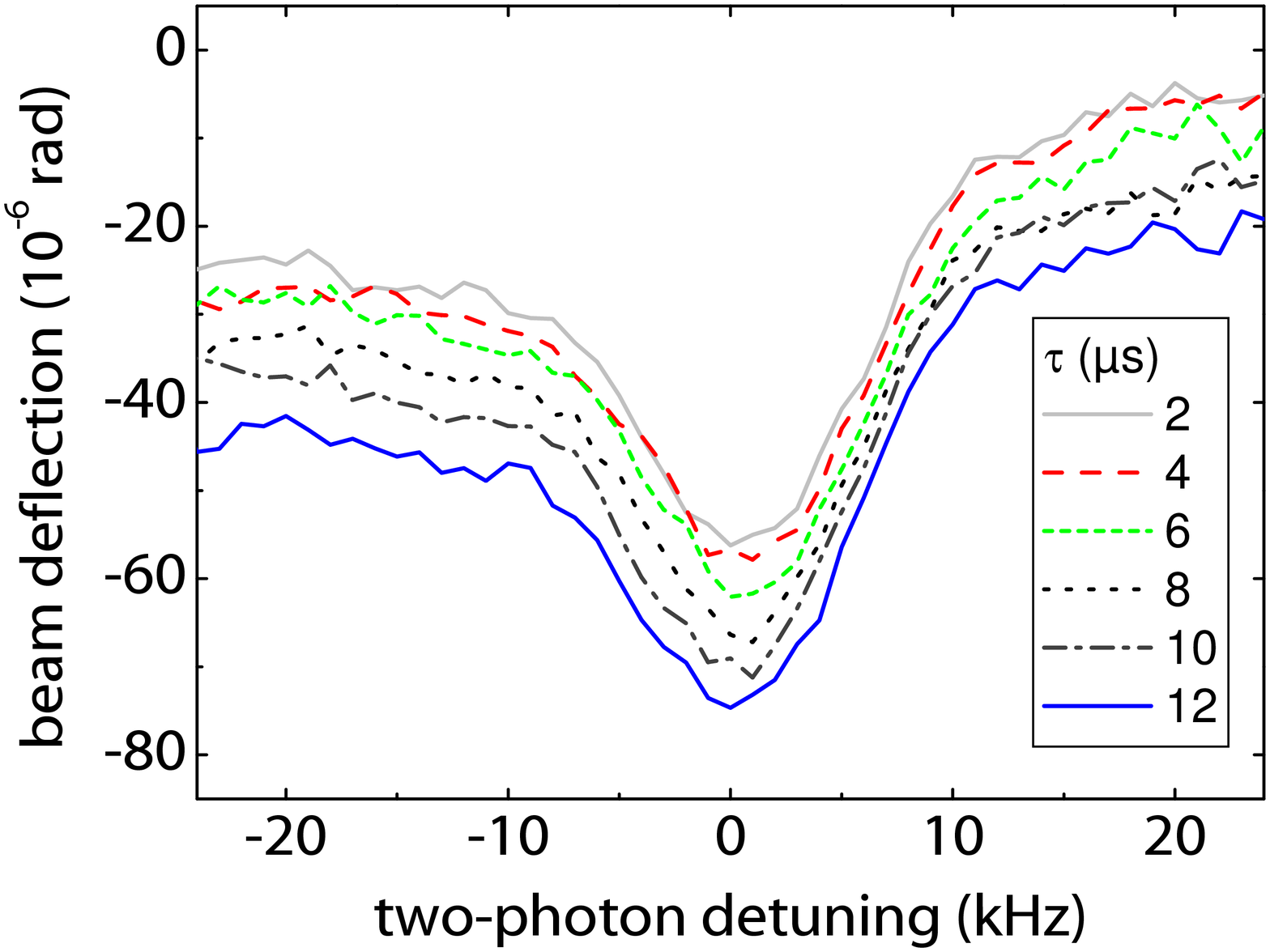}}
     \caption{\textbf{(a)} Relative signal beam angle deflection for different values of the light storage duration
     for the case of a two-photon detuning $\delta\cong 0$. The data has been shifted by the offset deflection acquired during the (non-stationary) phases of slow propagation to account solely for the deflection during the storage period.
           \textbf{(b)} Deflection of the retrieved signal beam versus the two-photon detuning for different values of the storage duration.}
     \label{fig:results}
\end{figure}
Fig. \ref{fig:figure4a} shows the obtained relative angle
deflection as a function of the storage time recorded with this
configuration. At our maximum available storage time of 20 $\mu
s$, the obtained beam deflection reaches $3\times10^{-5}$ rad. We
attribute an observed increase in uncertainty of the data for long
storage times to the here relatively small power of the retrieved
signal light, which increases the uncertainty in determining the
relatively small position variations after the short focal length
lens used in this experimental configuration. The data points have
been fitted with a linear function, yielding a slope of the
deflection angle versus the storage time of (1.30 $\pm$ 0.09)
rad/s. Within the quoted experimental accuracy, this value is in
good agreement with the result calculated from Eq.
\ref{eqn:AlphaStore}, which predicts a value (1.35 $\pm$ 0.08)
rad/s. The error bar of the expected value is dominated by the
experimental uncertainty in the determination of the magnetic
field gradient (described above). It is noteworthy that in case
that the polaritons would behave as massive particles that could
be moved during light storage by the applied field gradient, a
transverse beam displacement increasing quadratically with time
would be expected. Both a transverse displacement and a quadratic
dependence are inconsistent with the experimental data, which
shows an angular deflection that increases linearly with storage
time, in agreement with our theoretical model.

From our model, we also expect the beam deflection accumulated in
the stored light phase to be independent of the optical group
velocity. As the principal experimental result of this Letter,
Fig. \ref{fig:figure4b} shows the observed Stern-Gerlach
deflection for different storage times as the function of the
two-photon detuning $\delta$. For a given storage time $\tau$, one
observes a clear variation of the deflection on the two-photon
detuning. We attribute this dependence to deflection contributions
acquired during the propagating light phases of the storage
procedure, i.e. at times when the optical control field intensity
is different from zero during beginning and readout of the storage
sequence (see Fig. \ref{fig:figure2b}). The Stern-Gerlach
deflection acquired for a propagating polariton is clearly
dispersive and depends on the optical group velocity (see Eq.
\ref{eqn:AlphaMove}), so that the observed change in deflection is
well understood as due to the variation in optical group velocity
when sweeping over the two-photon resonance. When the storage time
is varied - the figure shows data recorded in incremental temporal
steps of 2 $\mu s$ - the experimental spectra exhibit a basically
equidistant displacement towards larger deflections. Noticeably,
the increase for larger storage times appears to be equidistant
for all shown values of the two-photon detuning, yielding evidence
for a non-dispersive deflection contribution acquired during the
storage phase. A cross section at $\delta=0$ corresponds to the
situation shown in Fig. \ref{fig:figure4a}. The observed angle
deflection can be well described as the sum of two contributions:
(i) a dispersive Stern-Gerlach deflection originating from the
interaction of moving quasiparticles of non-zero effective
magnetic moment with the magnetic field gradient and (ii) a
non-dispersive contribution arising from the phase gradient
imprinted during light storage. We point out that the observed
non-dispersive deflection effect also allows for the sensitive
detection of magnetic field gradients by evaluating the beam
position of the retrieved signal pulse. The obtained sensitivity
increases linearly with the storage time, and the
non-dispersiveness of the deflection represents a clear advantage
over techniques based on propagating optical fields
\cite{2006NatPh.2.332K}.

To conclude, we have investigated the deflection of an optical
beam by phase-imprinting by a Stern-Gerlach magnetic field during
the storage of light. We find that the dispersion is efficiently
suppressed with respect to that obtained in a Stern-Gerlach
deflection experiment of propagating atom-light polaritons,
realizing a proof-of-principle demonstration of non-dispersive
optics enabled by the dynamic variation of the optical group
velocity in electromagnetically induced transparency.

We expect that the observed effects allow for applications in the
field of coherent image storage and processing with
electromagnetically induced transparency. Related examples include
the development of all-optical deflectors or adaptive optics
components. It is an intriguing question, whether with advances in
the field of solid-state electromagnetically induced transparency
\cite{PhysRevLett.95.063601}, optical elements as e.g. lenses can be developed with the
dynamical variation of group velocity allowing for chromatic
aberration-free imaging. We also anticipate applications in
quantum memories, where the obtained deflection may allow for the
addressing of parallel channels \cite{Chaneliere2005,10.1038/nature04327}.
A different perspective includes applications in the field of
magnetic field metrology. The described techniques may be used to
characterize inhomogeneous magnetic field distributions from the
far-field deflection pattern of the retrieved signal field after light storage.

We thank W. Ketterle and F. Vewinger for helpful discussions.
Financial support from the Deutsche Forschungsgemeinschaft is
acknowledged.

\end{document}